# Towards Energy Efficient RAN:
# From Industry Standards to Trending Practice

Lopamudra Kundu, Xingqin Lin and Rajesh Gadiyar

NVIDIA

Email: {lkundu, xingqinl, rgadiyar}@nvidia.com

*Abstract*—As 5G deployments continue throughout the world, concerns regarding its energy consumption have gained significant traction. This article focuses on radio access networks (RANs) which account for a major portion of the network energy use. Firstly, we introduce the state-of-the-art 3GPP and O-RAN standardization work on enhancing RAN energy efficiency. Then we highlight three unique ways for enabling energy optimization in telecommunication networks, including full stack acceleration, network functions consolidation, and shared infrastructure between communication and artificial intelligence. These network design strategies not only allow for considerable overall reduction in the energy footprint, but also deliver several added benefits including improved throughput, reduced cost of ownership, and increased revenue opportunities for telcos.

## I. INTRODUCTION

The large-scale deployment of the fifth-generation (5G) mobile networks around the world has remarkably improved connectivity with unprecedented speed, low latency, and enhanced capacity. Nonetheless, there has been a growing concern on the operational costs and carbon footprint associated with the energy consumption of 5G networks. According to a recent benchmarking study by the global system for mobile communications association (GSMA), on an average 90% of energy use for an operator comes from the network, and the radio access network (RAN) accounts for more than 80% of the network energy use [1]. Optimizing the energy efficiency of the RAN is paramount for the telecommunications ecosystem, as the industry moves forward from 5G to the sixth-generation (6G) mobile networks.

Figure 1 provides an illustration of energy flow and main power consumptions in a typical RAN site. The site infrastructure consumes energy mainly due to air conditioning for cooling, rectifiers for converting alternating current (AC) to direct current (DC), and batteries for power backup. The main equipment, i.e., base station (BS) consumes energy mainly due to baseband unit, and radio unit. According to the statistics reported by the next generation mobile networks (NGMN) alliance, the BS is the largest power consumption component that consumes about 50% of the total power [2].

Various RAN power consumption models have been proposed in the literature to enable the evaluation of RAN energy efficiency as well as the assessment of different energy saving technologies. On a high level, the power consumption of a BS can be split into two parts: 1) dynamic part which is consumed for active data transmission/reception, and 2) static part which is consumed all the time even when the BS is idle. In the widely used RAN power consumption model presented in [3], the total power consumption at maximum load is proportional to the sum of power consumptions attributed by power amplifier (PA), baseband unit, and radio frequency (RF). The work [4] proposed a simplified BS power consumption model which considers massive multiple-input-multiple-output (MIMO) and ultra-lean design in 5G, where the BS power consumption increases linearly with the number of RF chains. More recently, the work [5] proposed a data-driven power consumption model for 5G active antenna units (AAUs), which considers multi-carrier PA technology besides massive MIMO. We refer interested readers to [6] for a detailed review of the various RAN power consumption models.

As we navigate the intricate landscape of energy efficiency in RAN, it is critical to grasp the latest industry developments, including global standardization efforts. Indeed, a variety of activities on RAN energy efficiency have been taking place in several standardization bodies, including the third-generation partnership project (3GPP) and the open RAN (O-RAN) Alliance, among others. This article comprehensively introduces the various technologies studied and standardized in 3GPP and O-RAN to enhance energy efficiency of RAN in Sections II and III, respectively.

While the energy saving mechanisms studied by 3GPP and O-RAN have led to standardization efforts towards making RAN more energy efficient at various levels, emerging implementation specific solutions are constantly pushing the boundary of energy savings envelope even further. Specifically, deploying network applications with software-defined, general-purpose accelerated platforms are poised to enable significant reduction in the telco energy footprint, while delivering better utilization and monetization of the network infrastructure. Accordingly, improving energy efficiency in RAN requires a holistic approach beyond narrowly focusing only on the RAN aspects. To that end, in Section IV, we highlight three unique ways to accomplish energy optimization in telecommunication networks on top of standardized energy saving best practices. These are: 1) full stack acceleration on a software-defined general-purpose platform, 2) consolidation of network functions on the same platform, and 3) co-hosting 5G and artificial intelligence (AI) applications on a unified infrastructure. Finally, the article wraps up in Section V with conclusion and scopes for future work.

## II. NETWORK ENERGY EFFICIENCY IN 5G-ADVANCED

In this section, we provide an overview of the state-of-the-art 3GPP work on network energy efficiency, spanning across

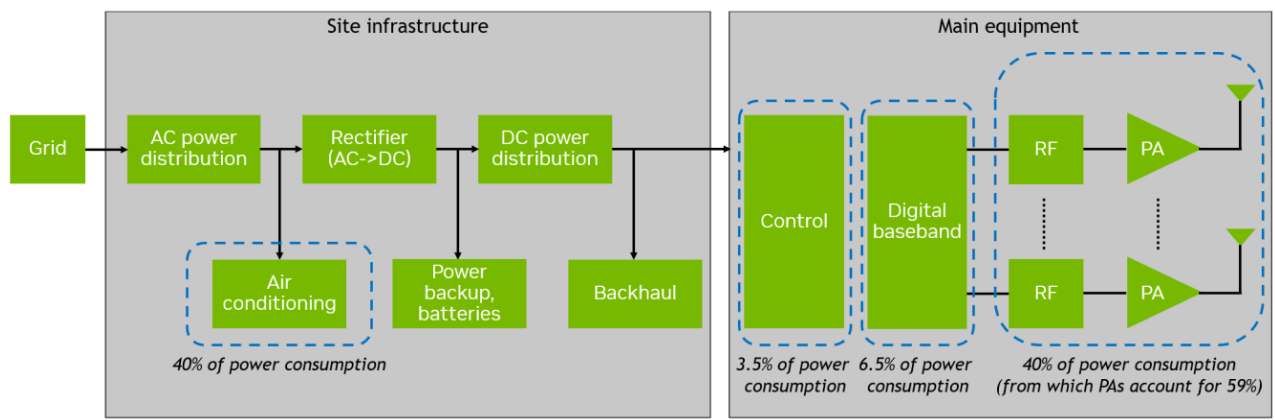

**Figure 1 An illustration of energy flow and main power consumptions in a typical RAN site. Indicative power consumption numbers are based on the NGMN report [1].**

multiple 3GPP groups from services and system aspects (SA) to RAN.

*A. 3GPP Energy Efficiency KPIs*

There are multiple ways to measure energy efficiency. The underlying basic idea of its key performance indicators (KPIs) is simple: *How much output can be delivered with one unit of energy?* The output can take various forms, e.g., data volume, connection number, and revenue. While each measure has its own advantages and disadvantages, the most widely used energy efficiency KPI for RAN is bit/Joule (J), defined as the ratio between the delivered data volume and the associated energy consumption. Energy efficiency KPIs defined for telecommunication networks can be applied to an entire end-to-end network, or a sub-network such as RAN, or a single network element, or a specific site, as specified in the technical specification (TS) 28.554 [7].

The 5G RAN (aka. next-generation RAN (NG-RAN)) data energy efficiency KPI (bit/J) is equal to the ratio between data volume and energy consumption of the considered network elements [8]. Besides the KPI calculation, interpreting variations in its values between different RANs may require additional parameters. These parameters describe the network characteristics such as population density, geographical conditions, and climate zones [9].

*B. 3GPP Study on Network Energy Savings for NR*

Network energy efficiency was a key consideration when NR was designed in its first release – Release 15, by minimizing the always-on signals, e.g., extending synchronization signal periodicity and removing cell-specific reference signal (CRS) present in long-term evolution (LTE) [10]. As commercial 5G deployments are expanding throughout the world, 5G network energy saving is becoming even more critical for environmental sustainability and operational expenditure (OpEx) reduction. These considerations motivated 3GPP RAN working groups to conduct a study on network energy savings for NR in Release 18 [11].

As part of the study, 3GPP developed a BS energy consumption model. The model is based on three sets of reference configurations that differ in duplex mode, system bandwidth, subcarrier spacing, number of transceiver chains, and transmit power. The model features five power states with different relative power consumption values: deep sleep, light sleep, micro sleep, active downlink, and active uplink. In the deep sleep state, minimal power is consumed with very few components (e.g., internal clocks) on and a long transition time of the order of several tens of milliseconds or more. In the light sleep state, many components such as PA, low noise amplifier (LNA), and integrated circuits in transmitter and receiver units can be turned off, leading to low power consumption and a transition time of at least several milliseconds. In the micro sleep state, components (e.g., PA and LNA) with a transition time of the order of microseconds can be turned off to reduce power. The two states of active downlink and active uplink are used to model BS power consumption for downlink and uplink separately. In these two active states, all components are switched on and thus they consume the highest power among the five power states.

With the agreed BS energy consumption model, 3GPP investigated a comprehensive set of network energy saving techniques in time, frequency, spatial and power domains, as illustrated in Figure 2 and summarized below:

- *Time domain techniques* aim to create more idle opportunities in time for the 5G node B (gNB) so that it can enter sleep modes more often to save energy. As an example, adapting the transmission/reception patterns of common channels or signals, such as increasing the periodicity and skipping certain transmissions/receptions, can provide more sleep opportunities for the gNB.
- *Frequency domain techniques* adapt the BS's operation in frequency domain, such as turning off carriers, reducing operating bandwidth, or decreasing transmission/reception of certain channels or signals within a carrier, to save energy. As an example, a secondary cell (SCell) in carrier aggregation (CA) may be operated without synchronization signal block (SSB) for network energy savings.
- *Spatial domain techniques* refer to the methods that change the number of used antennas at different levels, such as activation/deactivation of cells, transmission reception points (TRPs) of a cell, antenna panels of a TRP, and antenna elements of a panel. As an example, when the traffic load is low, a subset of transceivers in a massive MIMO BS can be muted to save energy.

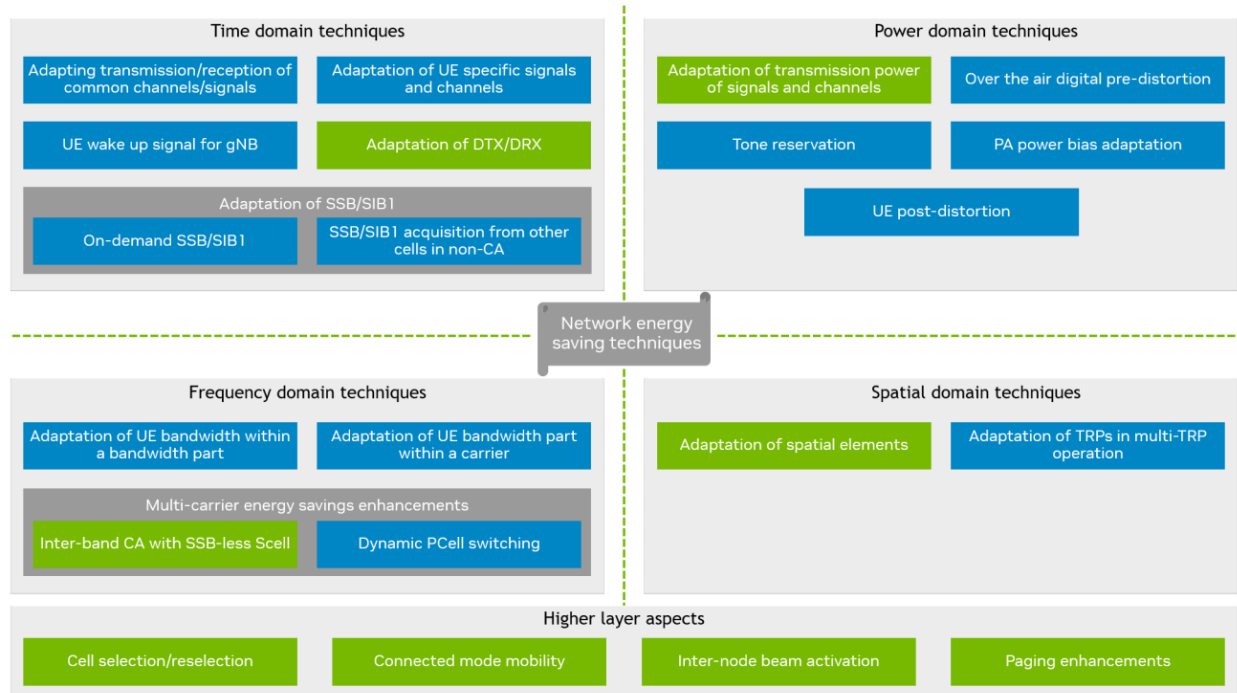

**Figure 2** An overview of network energy saving techniques studied by 3GPP in Release 18. Techniques in blue and green boxes were investigated in Release-18 study item and those techniques in green boxes were further included in Release-18 work item.

- *Power domain techniques* encompass methods to adapt the transmission power of channels or signals and approaches to improve PA efficiency. As an example, enabling PA operation in the non-linear region with digital pre-distortion and received signal post-distortion can improve PA efficiency and thus save network energy.

While various techniques can be used for network energy savings, there is a tradeoff between saving network energy and impacting network and user performance. That is why, for a given technique, 3GPP study not only evaluated potential network energy saving techniques, but also examined their impact on several other KPIs such as spectral efficiency, capacity, latency, and user equipment (UE) power consumption, among others.

*C. 3GPP Enhancements on Network Energy Savings for NR*

With the completion of the Release-18 study on network energy savings for NR, 3GPP is working on the normative work to specify enhancements in Release 18.

In the *time domain*, a gNB can reduce its downlink transmission and uplink reception activities by using cell discontinuous transmission (DTX) and discontinuous reception (DRX) mechanisms. To this end, 3GPP specifies enhancements to enable inter-node information exchange about cell DTX/DRX as well as to align cell DTX/DRX and UE connected mode DRX.

In the *frequency domain*, operation without SSB (i.e., SSB-less) for an SCell in inter-band CA is a promising network energy saving technique. To this end, a UE can measure SSB transmitted on a primary cell (PCell) or a different SCell with SSB to enable time and frequency synchronization on the SSB-less SCell as well as layer 1/layer 3 measurements.

In the *spatial domain*, adaptation of spatial elements such as antenna ports and active transceiver chains has been identified to be an effective network energy saving technique. In this area, 3GPP specifies necessary enhancements on channel state information (CSI) and beam management related procedures including signaling, measurement, and reporting to enable efficient adaptation of spatial elements.

In the *power domain*, the gNB can adapt its physical downlink shared channel (PDSCH) transmit power to save network energy. To this end, signaling enhancement for providing updates on the power difference between the PDSCH and CSI reference signal (CSI-RS) is necessary, because UE receives CSI-RS and calculates CSI by considering the power difference of the two transmissions.

In addition, 3GPP specifies enhancements on inter-node beam activation, restricting paging in a limited area, cell selection, and conditional handover which will take into account whether the source or target cell is in network energy saving mode. It is also worth noting that in a separate work item on AI/machine learning (ML) for NG-RAN in Release 18, 3GPP specifies data collection enhancements and signaling support for AI/ML-based network energy savings [12].

Overall, the energy saving strategies standardized by 3GPP are focused on enhancing RAN signaling, configurations, and protocols. While these enhancements lay the foundation for energy-efficient network design, the latest trend of RAN transformation towards openness and virtualization brings new opportunities with additional mechanisms and tools that can be augmented on top of energy saving approaches guided by 3GPP [13]. To that end, the following section introduces open RAN centric energy saving techniques that are studied and specified by O-RAN Alliance.

III. ENERGY EFFICIENCY IN OPEN RAN

In recent years, the evolutionary journey of 5G networks towards open RAN has gained significant momentum. The core principle of open RAN centers around disaggregating single-vendor, monolithic RAN architecture into multi-vendor, modular components, interoperable through open and standardized interfaces. Making large scale deployment of open RAN a reality requires industry-wide standardization effort led by O-RAN Alliance.

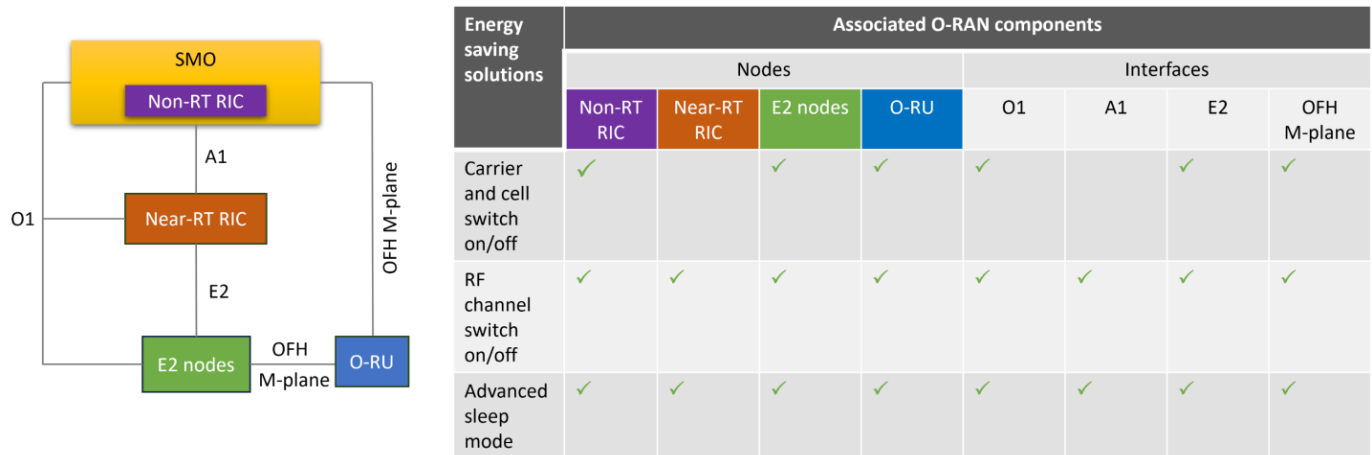

Figure 3 An overview of O-RAN architecture and its associated nodes/interfaces enabling energy saving solutions.

O-RAN Alliance has augmented the 3GPP-defined NG-RAN architecture with disaggregated nodes, open interfaces, and most importantly, RAN intelligent controller (RIC) – a newly introduced network function with native AI/ML capability. RIC manifests in two different timescales: non-real time (non-RT) RIC and near-real time (near-RT) RIC. RIC nodes offer new opportunities for designing agile, automated, and highly efficient open RAN. In the context of RAN efficiency, the measurement metric can have many dimensions, one of which is energy efficiency.

Energy saving mechanisms studied by O-RAN Alliance working group 1 (WG1) primarily focus on software-based solutions aided by AI/ML at an architectural level, which can be augmented on top of 3GPP-defined techniques [13]. These softwarized solutions aim at reducing energy consumption by optimizing configurations of O-RAN network elements and control signaling to improve the overall energy rating of the network. Energy savings can be enabled at different timescales, which determines the set of 'knobs' that can be tuned within the O-RAN architecture. To that end, the energy saving mechanisms studied by WG1 are mapped into three categories depending on the solution timescale: non-RT, near-RT, and real time (RT) [14][15]. Figure 3 shows an overview of O-RAN architecture with its key nodes/interfaces associated with enabling various energy saving solutions. Each of these solutions is detailed in the following subsections.

### A. Carrier and Cell Switch On/off in Non-real Time

O-RAN has leveraged its support of native AI/ML services and standardized open interfaces to enable cell/carrier switching. When the cell traffic load is low, O-RAN enables switching off one or more carriers, or even an entire cell, to reduce the power consumption of various components, such as open–central unit (O-CU), open-distributed unit (O-DU) and open-radio unit (O-RU). There are two modes of operation: 1) hibernation mode (wherein O-RU's PAs remain in a state with minimum current rating) and 2) complete switch-off mode (wherein the PAs are entirely shut off).

Figure 4 shows a call flow diagram of carrier and cell switch on/off enablement in O-RAN architecture. At a timescale of the order of minutes and above (i.e., non-RT), the network element parameters related to carrier and cell switching are configured by non-RT RIC via open interfaces, viz. O1 interface (for O-CU/O-DU, known as E2 nodes) and open fronthaul management plane (OFH M-plane) interface (for O-RU). While non-RT RIC collects network traffic information at a per-cell level and/or per-carrier level, the applications running on the non-RT RIC platform (also known as rApps) use this information as an input to provide AI/ML-assisted solutions. Essentially, these solutions derive optimal energy saving configurations for the network elements, for example, cell and carrier on/off switching time and traffic steering strategy that can achieve the intended tradeoff between energy consumption and users' guaranteed quality of service (QoS). The deployed AI/ML models at non-RT RIC can include functionalities like prediction of future cell traffic, user mobility, network resource usage and expected energy savings for different optimization configurations, among others. As per the derived energy saving policy, non-RT RIC implements corresponding configurations for carrier/cell switching and traffic steering.

O-RAN has conducted gain analysis for this energy saving scheme, focusing on a couple of candidate deployment scenarios [15]. For example, with O-RU configurations including 4-transmit-4-receive (4T4R) antennas, 4 layers, 100 MHz bandwidth, 3.5 GHz carrier frequency and 30 Watt of transmit power per antenna, the power saving during carrier/cell switch off (low load) can potentially be up to 150 ~ 180 Watt/O-RU/hour. With the O-RU configurations upscaling to 64T64R and 16 layers, the power saving can go up to 260 ~ 340 Watt/O-RU/hour. Details for corresponding O-RU power profiles are reported in [15].

### B. RF Channel Switch On/off in Near-real Time

At a timescale of the order of seconds (i.e., near-RT), O-RAN enables energy saving mechanisms through switching on/off RF channels and independently, switching on/off transmitter and receiver chains. These energy saving techniques are especially useful for open RAN networks with massive MIMO deployment. Unlike carrier and cell switch on/off that is typically applied at a low cell traffic load, RF channel switch on/off can be used for energy saving even when the cell traffic load is moderate.

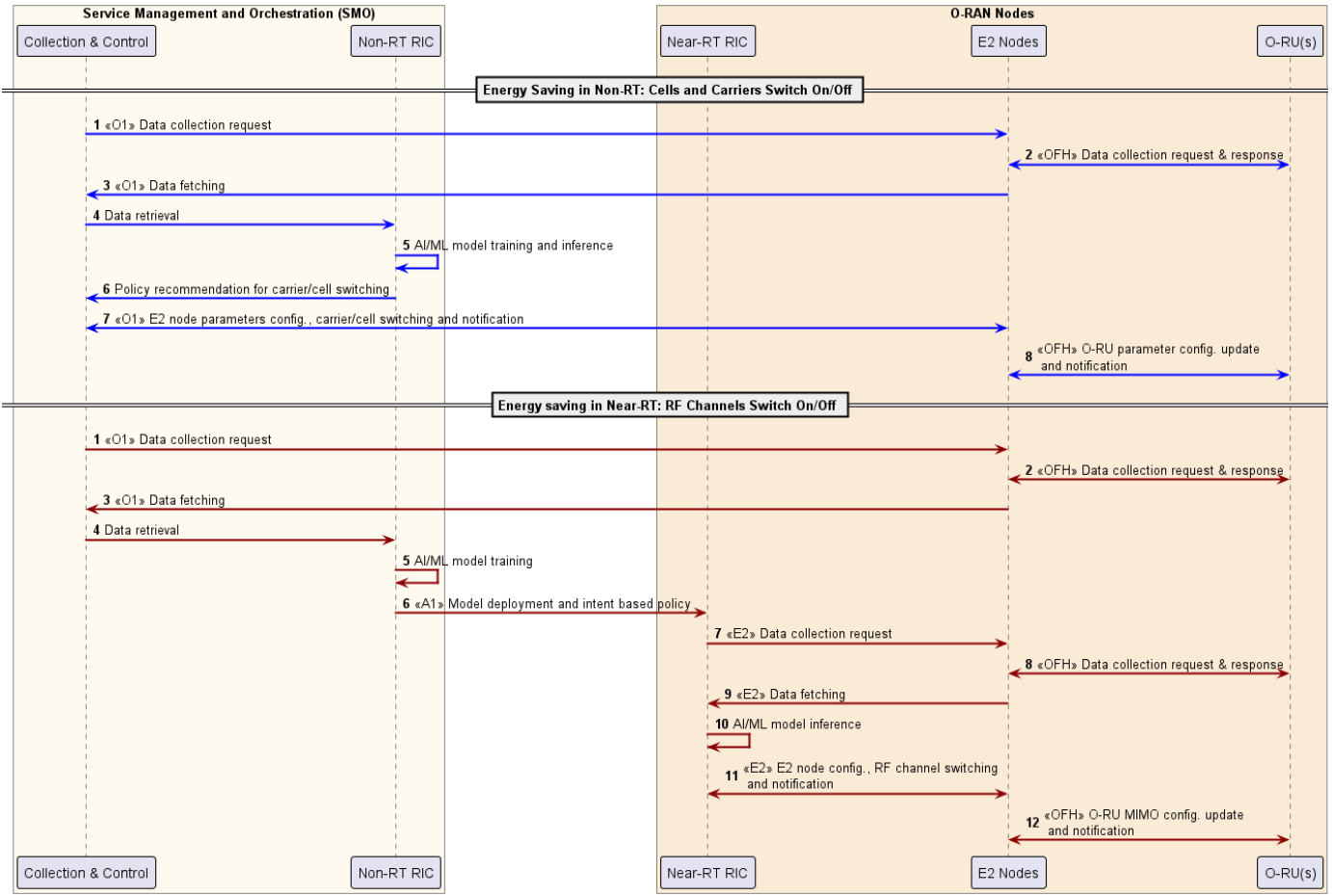

**Figure 4** Flow diagram for O-RAN energy saving mechanisms implemented at two timescales: non-RT and near-RT.

Figure 4 shows a call flow diagram of RF channel switch on/off enablement in O-RAN architecture. Based on the measurements collected from O-CU/O-DU/O-RU, the energy saving techniques assisted by AI/ML automatically switch on/off certain RF channels and deploys massive MIMO system configurations that minimize impact on users' QoS while optimizing energy savings. Relevant xApps running on near-RT RIC trigger RF channel switching considering traffic information such as cell load, user location, and user mobility. While switching off RF transceivers and/or antenna elements is used for digital beamforming based massive MIMO, reducing number of layers and/or multiuser scheduled UEs is used for hybrid beamforming based massive MIMO to save energy.

Enabling RF channel switching in near-RT involves non-RT RIC as the AI/ML model training host, and near-RT RIC as the AI/ML model inferencing host. Model training is performed based on data collected over O1 interface. Subsequently, near-RT RIC uses the trained model to infer RF channels' switching time as well as optimum massive MIMO system configurations. Corresponding O-RU reconfigurations are applied over OFH M-plane interface.

Gain analysis conducted by O-RAN for similar deployment scenarios as mentioned in the previous subsection shows that relatively less power saving is achievable with RF channel switching compared to carrier/cell switching given the same O-RU configurations and power profiles, e.g., ~ 80 Watt/O-RU/hour for 4T4R and ~230 Watt/O-RU/hour for 64T64R [15].

### C. Advanced Sleep Mode in Real Time

Energy saving at a very short timescale (of the order of milliseconds, i.e., frame/subframe/symbol duration) is enabled through advanced sleep mode (ASM), wherein O-RU components are partially switched-off. ASM can be invoked even when the cell traffic load is high. Configurations of cell-specific parameters (e.g., SSB periodicity) needed for operations of ASMs are derived automatically by near-RT RIC using multifaceted data including cell traffic load, energy efficiency KPI measurements, and user service types, among others.

Switch-off time duration and the transition to switch-on time are interrelated, implying that the longer the switch-off time or sleep mode, the longer it takes to transition to the switch-on time or wake-up mode. While aiming at maximizing ASM duration, the energy saving strategies take into consideration other user specific constraints (e.g., latency) and accordingly, derive optimum ASM activation/deactivation policy and configurations to achieve the desired tradeoff between energy savings and QoS. System configurations associated with ASM policies can be applied to various instances, for example, setting SSB periodicity for certain frames, compressing data transmissions in specific slots, preempting slots that are designated to sleep state, reducing transmission of synchronization signals, and shrinking opportunities for random access and paging.

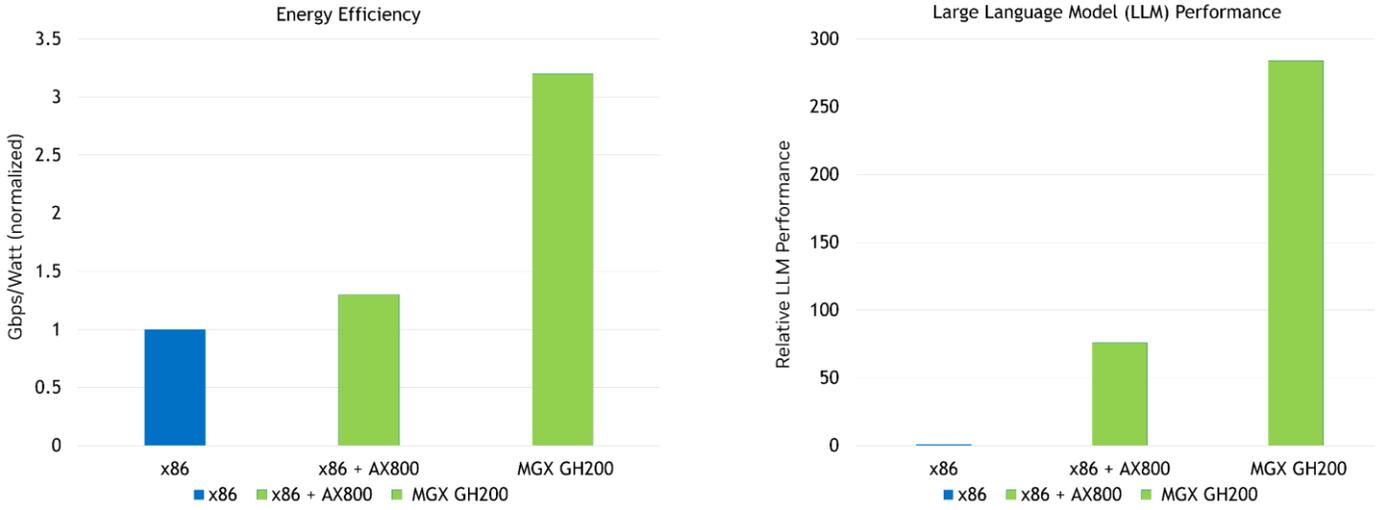

**Figure 5** Comparison of throughput, energy efficiency and LLM performance between CPU and GPU accelerated compute.

To automate the derivation of ASM policies/configurations, AI/ML-assisted mechanism is used, wherein a split training-inferencing approach is adopted using O-RAN's non- and near-RT RIC frameworks, similar to the RF channel switch on/off call flow depicted in Figure 4.

In addition to these energy saving modes, O-RAN is also looking into potential solutions to reduce energy consumption at node levels of open cloud (O-cloud), including adaptive hardware shutdown, scaling out network functions, and optimizing power, memory, and frequency of general-purpose as well as special-purpose compute elements within O-cloud.

## IV. ENERGY SAVING PRACTICE BEYOND STANDARD

In this section, we explore various emerging trends beyond standardization that are compelling for further energy efficiency improvement as the industry marches towards 6G. Industry progress in cloudification, distributed computing and accelerated computing are key enablers for these end-to-end optimizations across 5G RAN and 5G Core (5GC). We highlight three unique network design strategies which not only allow for considerable overall reduction in the energy footprint, but also deliver several added benefits, described in the following subsections.

### A. Full Stack Acceleration of Network Functions

There are significant opportunities for workload acceleration in both 5G RAN and 5GC network functions. Within RAN domain, workload acceleration opportunities include but not limited to 1) software-defined, end-to-end inline acceleration of layer 1 of virtual RAN (vRAN), 2) acceleration of layer 2 functions, e.g., medium access control (MAC) scheduler acceleration, multi-cell scheduling with interference avoidance, 3) AI-aided/enabled signal processing including advanced beamforming techniques for massive MIMO, among others.

In 5GC, the data plane functions associated with the user plane function (UPF) such as general packet radio service (GPRS) tunneling protocol (GTP), packet classification, receive side scaling (RSS) hashing and security can be accelerated using smart network interface card (SmartNIC) or data processing unit (DPU). Furthermore, the UPF can be distributed and deployed closer to the edge as intermediate UPF for local break out (LBO) of data traffic associated with the edge applications such as generative AI (GenAI) to improve latency and QoS.

Acceleration of compute-intensive workload processing improves the performance of the network functions within the same power envelope, leading to better energy efficiency of the overall network. Figure 5 shows a comparison between three compute platforms which are based on x86 (central processing unit (CPU) only), x86 with graphics processing unit (GPU) acceleration boost by NVIDIA AX800 converged card, and NVIDIA MGX GH200 converged platform with ARM CPU. The performance KPIs are scaled with respect to 'x86 only' compute. It is evident from Figure 5 that there is a pronounced performance benefit in terms of vRAN performance (e.g., throughput and energy efficiency), and large language model (LLM) performance by utilizing accelerated computing platforms for both 5G vRAN and AI applications.

### B. Network Functions Consolidation

Over the last decade, the telecommunication industry has widely embraced the notion of network virtualization and cloudification. There is a growing trend towards using Kubernetes and cloud native deployment of network functions.

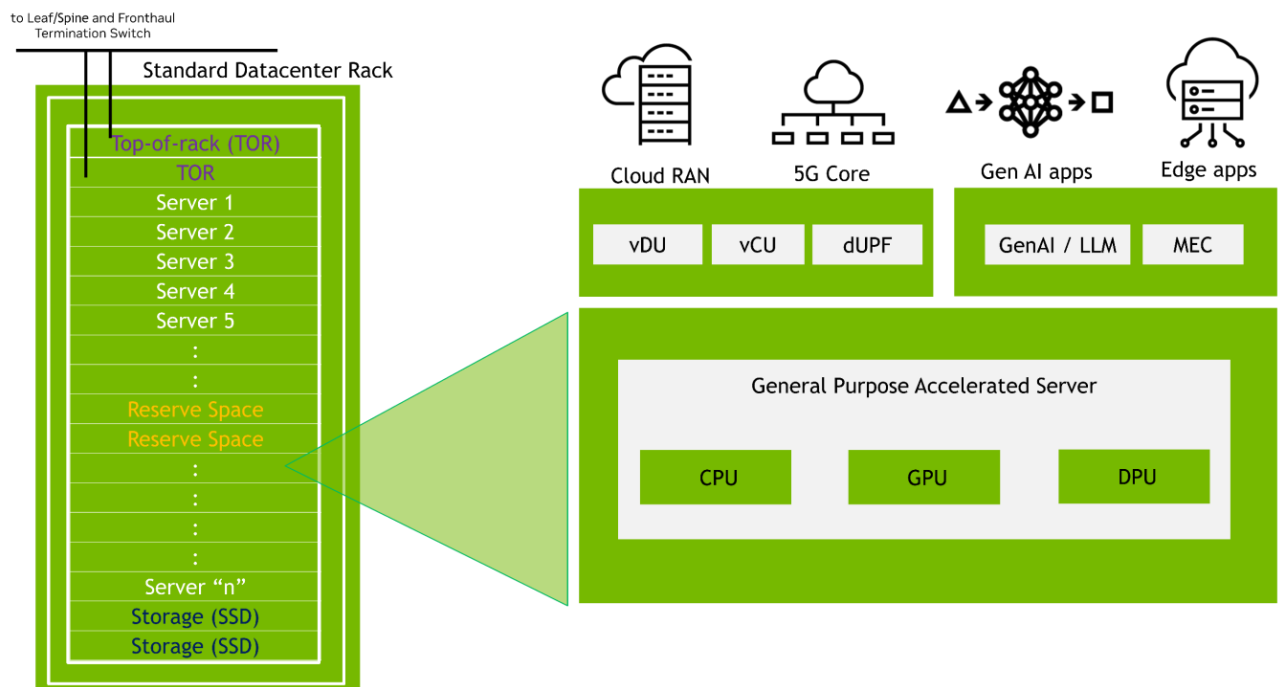

Figure 6 Shared AI and telecommunication infrastructure with software-defined distributed edge cloud.

This has made orchestration of virtualized network functions more agile and scalable. The next frontiers emerging in the advanced network management space are intent-based closed loop automation, and dynamic orchestration and consolidation of network functions, aiming to drive improved utilization of the infrastructure and reduced energy footprint of telecommunication networks.

In a cloud deployment, it is possible to run a plethora of heterogeneous workloads comprising vRAN network functions (e.g., virtual DU (vDU)/virtual CU (vCU)), distributed UPF, virtual router and security functions (e.g., firewall and virtual private network (VPN) gateways) in the same cloud datacenter or even the same server rack. It is well known that server platforms that are not fully utilized result in higher average energy consumption. As an illustrative example, a server that is 50% utilized on an average would consume 150 Watt, while the same server, if 90% utilized, may consume 185 Watt. As a result, it is suboptimal when a vRAN/5G datacenter uses many CPU servers that on the whole run at a low utilization as compared to a densely packed accelerated datacenter with small number of highly utilized accelerated computing servers such as NVIDIA MGX GH200. A key requirement to accomplish this consolidated network functions/applications hosting, and efficient processing is a fully software-defined architecture blueprint of cloud infrastructure deployed on general-purpose hardware platform, enabling acceleration of key network functions. Aside from the lower latency benefit of co-location, dense packing of network functions on a homogeneous cloud platform results in much higher resource utilization and net gain in energy savings. Figure 6 shows a software-defined accelerated datacenter with general-purpose commercial-off-the-shelf (COTS) servers that are built with a unique CPU, GPU and DPU based architecture. The network functions in the RAN and 5GC are consolidated on this accelerated platform. Additionally, the platform can also run GenAI and LLM applications sharing the same infrastructure and unlock significant monetization opportunities.

### C. Shared AI and Telecommunication Infrastructure

Traditionally, telcos build dedicated infrastructures for different parts of their networks, providing different services. In that practice, RAN has been primarily deployed with purpose-built platforms. More recently, the industry has started pivoting towards open RAN, where network function softwarization, cloud-native methodologies, and deployment with COTS servers are replacing the vendor-specific, single-purpose platforms. Up until now, the high-performance demand for 5G has been addressed by the adoption of fixed-function accelerators, to supplement the compute horsepower of general-purpose processors like CPU. 5GC network functions, on the other hand, had started this virtualization journey a while ago, with most of the commercial deployments today being on standard, COTS-based servers.

With the advent of GenAI, the telecommunication industry has seen an unprecedented growth in compute demand for running multitudes of LLM-based applications at the edge, resulting in massive growth of edge GPU servers for executing LLM inferencing and other multi-access edge compute (MEC) applications. As the heterogeneity of applications running on the same network infrastructure is proliferating rapidly, it is fueling multiple isolated clusters within the same network dedicated for running various parts and applications of the network. With the highly dynamic and time-varying nature of network traffic, not all these clusters' hardware is fully utilized all the time, leading to gross underutilization of the compute resources throughout every part of the network infrastructure and, as a side effect, an overall increase in the network energy footprint.

One way to resolve this impending suboptimality of deployed network would be to build an overarching, homogeneous, multi-tenant and software-defined cloud infrastructure that can run both cloud RAN and core network functions, while at the same time, can also execute the AI and MEC applications by sharing the compute resources of the underlying platform. Dynamic orchestration of AI and network function workloads on the same infrastructure would allow for optimum resource utilization, with softwarized resources repurposed and shared across multiple workloads as per their compute needs. Figure 6 depicts a high-level architecture of a shared AI and telecommunications infrastructure that can be built on a standard datacenter rack.

To summarize, the unique network design approaches discussed above offer three-fold added benefits on top of energy savings: 1) increased performance with network function acceleration, 2) reduced total cost of ownership by consolidating multiple network functions on the same infrastructure, and 3) increased monetization by deploying GenAI, especially LLM applications, on a shared, distributed edge infrastructure.

## V. CONCLUSION AND FUTURE WORK

As the industry moves forward from 5G to 6G, improving RAN energy efficiency is critical for sustainability. Throughout this article, we have explored RAN energy saving techniques from industry standards to implementation practice. We introduced the state-of-the-art 3GPP and O-RAN standardization work on enhancing RAN energy efficiency. Furthermore, we highlighted three unique ways for energy optimization in telecommunication networks, including full stack acceleration, network function consolidation, and shared infrastructure between telecommunication and AI.

We conclude by pointing out some of the important topics for future work.

- *Joint computing and communication*: Computing has been increasingly integrated into communication infrastructure to support important use cases such as GenAI applications. It is paramount to investigate novel architectures that are inherently designed for energy efficiency to support joint computing and communication.
- *Intelligent network optimization*: Intelligent and automated network planning, management, and optimization are crucial for network energy efficiency. The application of AI/ML techniques to optimize network energy usage deserves further exploration.
- *Standardization*: While standards bodies like 3GPP and O-RAN have carried out work to enhance network energy efficiency, more needs to be done. In particular, 6G standardization should focus on energy efficiency right from the beginning and understand the key enablers for sustainable 6G.

ACKNOWLEDGEMENT

We acknowledge the important engineering work conducted by several teams at NVIDIA, which was instrumental in the derivation of performance benchmarks reported in Figure 5.